\documentclass{article}
\usepackage{frascatiphys}
\usepackage{graphicx}
\begin{document}
\title{ 
NEW PROJECTS ON DARK PHOTON SEARCH
}
\author{
Venelin Kozhuharov  \\
{\em University of Sofia, 5   J. Bourchier Blvd, 1164 Sofia, Bulgaria} \\
{\em  INFN - LNF, Via E. Fermi 40, 00044 Frascati (RM), Italy  } \\
Mauro Raggi \\
{\em  INFN - LNF, Via E. Fermi 40, 00044 Frascati (RM), Italy  } \\
Paolo Valente\\
{\em INFN - Sezione di Roma, P.le Aldo Moro, 2 - 00185 Roma - Italy}
}
\maketitle
\baselineskip=11.6pt
\begin{abstract}
Despite the great success of the Standard Model of particle physics
the nature of Dark Matter still remains unclear. 
Recently, the idea of the existence of a hidden sector coupling only
 weakly with the ordinary matter was revitalized and gained
 popularity.
 A simple mediator between the hidden and the visible sector could be
 a 
vector particle of a new gauge interaction, the so called dark
photon. 
Numerous of activities were initiated to probe its parameter space. 
The present results and the foreseen experiments aimed to search for 
dark photons in few directions are reviewed and discussed.

\end{abstract}
\baselineskip=14pt

\section{Introduction} 

The Standard Model (SM) of particle physics provides 
a unique description of almost all phenomena in the 
microworld. However, the unexplained nature of Dark Matter
indicates the necessity of its extension to a more fundamental 
theory, which incorporates also the cosmological observations. 
This can be realized in a large variety of theoretical models.

From experimental point of view the best path to follow is
to focus on few of the observed ``smoking guns'' indicating 
a possible discrepancy between the SM prediction and the results from the data. 
At present there are still unexplained anomalies  with 
the annually modulated excess of signal observed by DAMA/Libra \cite{bib:dama-libra};
the positron and antiproton excess in the cosmic rays 
\cite{bib:pamela2008,bib:fermi-pos,bib:ams-pos,bib:ams-antiproton}; 
the three sigma discrepancy in the 
anomalous magnetic moment of the muon \cite{bib:g-2-mu}; 
the failure to explain the $^8$Be anomaly \cite{bib:be-anomaly}
within nuclear physics. 
All those measurements may indicate the presence of new 
undiscovered so far particles.

%
%
%
%
%
%

An elegant model to address the possible existence of new 
degrees of freedom is the concept of a dark sector (DS) of 
particles interacting only weakly with the Standard Model particles.
The origin of the interactions could be due to the presence of a
mediator - an object carrying both SM and DS quantum numbers or SM 
fields possessing a relatively small charge under any of the gauge 
symmetries in the Dark Sector. In that sense the mediator 
provides a so-called portal to the DS \cite{bib:dark-sector-report-US} \cite{bib:tom-vulcano}. 
One of the simplest possible realization 
that is ultra violet safe 
(i.e. does not introduce new scale in the lagrangian) 
is by employing a vector gauge field, the so-called Dark Photon A' (DP), 
which interacts weakly with the SM fermions
\begin{equation}
 \mathcal{L} ~\sim ~ g' q_f \bar{\psi}_f\gamma^{\mu}\psi_f A'_{\mu},
\label{eq:dp-general}
\end{equation}
where $g'$ is the universal coupling constant and 
$q_f$ are the corresponding fermion charges. 
The term in equation (\ref{eq:dp-general}) could  be 
effectively realized also through kinetic mixing
 of the massive DP with the 
ordinary photon. 
In this picture the interaction of the SM particles with 
the dark photon will be described by two parameters - 
$\epsilon \sim g'q_f$ and the dark photon mass $m_{A'}$. 
The coupling parameter $\epsilon$ could be flavour dependent
giving rise to different 
leptophobic or leptophilic DP models \cite{bib:dpreview}.

Two different scenarios depending on the phenomenology in 
the dark sector could be identified. 
In the case when no new light degree of freedom $\chi$ exists in the dark sector 
  the dark photon 
will decay to SM particles only, with Br($A'\to e^+e^-$)~=~100\% for 
1~MeV~$< m_{A'} ~<$~210~MeV. 
If, however, $m_{\chi} < m_{A'}/2$ then the decay $A'\to\chi\chi$ will be 
dominant since it is not suppressed by the small value of $\epsilon$. 
In the latter scenario the observables will depend on two additional parameters, 
the coupling strength in the dark sector $\alpha_{D}$ and $m_{\chi}$.
These two scenarios, so called ``visible'' and ``invisible'',
result in very different experimental signature.
In the visible case a narrow resonant might be
observed in the dilepton or, in general, in the di-particle invariant 
mass spectrum, 
while in the invisible case the existence of a DP could
present itself through the search for  
missing mass or missing energy.

\section{Visible dark photon decays}

Currently, most of the experiments addressing the existence 
of an $A'$ performed ``peak'' searches in the $e^+e^-$ invariant 
mass spectrum. 
This approach requires a precise spectrometer providing measurement
of the electron momentum with high precision. The production of the $A'$ 
could be either through a bremsstrahlung process (in electron-on-target experiments)
or through the $e^+e^- \to \gamma A'$ at electron-positron colliders. 

Two new dedicated to the A' search 
experiments are planed in the near future - the HPS experiment 
at the JLaB \cite{Celentano:2014wya,bib:hps} and the MAGIX experiment at Mainz \cite{bib:magix}. 
Both of them will exploit an electron beam 
impinging on a thin target.

The HPS experiment will measure the momentum 
of A' products using a silicon vertex tracker 
placed inside a dipole magnet. 
A downstream lead tungstate calorimeter will serve for 
fast energy measurement and triggering. 
The silicon tracker is made of six dual sensor layers 
and will allow to address the dark photon decays in two different 
regions of its parameter space. 
For large values of $\epsilon^2$  ($\epsilon^2 > 10^{-7}$) 
A' decays promptly and the event selection will be based on the 
$e^+e^-$ invariant mass reconstruction. The dominant background
originates from the internal pair conversion of the bremsstrahlung photon
into $e^+e^-$ pair.
When $\epsilon^2 \leq 10^{-8}$ the finite lifetime of $A'$ will result 
in the reconstruction of vertices displaced from the interaction point inside 
the target. This channel is particularly interesting since 
it could be background free, at the expense of a low signal yield. 
The projected sensitivity for HPS is $\epsilon^2 < 3-4\times 10^{-7}$ for 
20~MeV~$\leq m_{A'}\leq$~300~MeV and 
$2\times10^{-8} \leq \epsilon^2 \leq 2\times 10^{-10}$ in the region 
20~MeV~$\leq m_{A'}\leq$~200~MeV

The MAGIX experiment is planned to operate 
at the new energy-recovering superconducting accelerator at Mainz, MESA,
which provides an electron beam with energy up to 155 MeV and
1 mA beam current. 
The target will be accomplished as a 
windowless supersonic gas jet 
with high density ($10^{19}/cm^2$). 
The A' decay products, $e^+$ and $e^-$, will be detected in a double arm 
spectrometer with planned resolution of $\delta p/p = 10^{-4}$. 
The projected sensitivity indicates reach of $\epsilon^2\sim10^{-8}$ for 
masses 10~MeV~$\leq m_{A'}\leq$~50~MeV.

\section{Invisible dark photon decays}

The mass spectrum of the particles in the 
dark sector is in general ambiguous and nothing
prevents the existence of light states. 
Such a scenario is relatively difficult to probe 
due to the impossibility to perform a complete 
reconstruction of the final state when A' decays
to $\chi\chi$. 
Thus it is important to start with an initial state
that can be fully described. The annihilation of 
$e^+$ with $e^-$ is one of the possibility while another 
is to usage of mono-energetic beam and to
search for missing energy taken away by the produced 
dark photons.

\subsection{Missing mass technique}
The missing mass technique relies on the complete 
reconstruction of the annihilation process 
\begin{equation}
 e^+ + e^- \to \gamma A'
\end{equation}
through the measurement of the energy and the direction of the recoil photon.
In positron on target collisions the missing mass squared is then computed as 
\begin{equation}
M^2_{miss} = (P_e + P_{beam}-P_{\gamma})^2
\end{equation}
where usually the electron is considered to be at rest ($P_e = (m_e,0,0,0)$). 
The cross section for the production of $A'$ is 
enhanced with respect to the ordinary $e^+e^-\to2\gamma$ process 
by a factor $\delta$ \cite{bib:padme-proposal}, especially for 
$m_{A'}$ close to the centre of mass energy of the interaction.
The dominant background processes are listed in table \ref{tab:mmiss-bkg} and
originate from the 
bremsstrahlung emission in the field of the target nuclei and from 
the  $e^+ + e^- \to \gamma \gamma \gamma$ annihilation. Usually, the measurement of 
the energy of the recoil photon provides  missing mass squared resolution good enough to
suppress the $e^+ + e^- \to \gamma \gamma$ background.

\begin{table}[t]
\centering
\caption{ \it Dominant background contributions to the missing mass technique}
\vskip 0.1 in
\begin{tabular}{|l|c|c|} \hline
Background process & $\sigma$ ($E_{beam}$ = 550 MeV) & Comment \\
\hline\hline
$e^+e^-\to\gamma\gamma$  & 1.55 mb &  \\
$e^+N \to e^+ N \gamma$ & 4000 mb & $E_{\gamma} > 1 MeV$, on carbon\\
$e^+e^-\to\gamma\gamma\gamma$  & 0.16 mb &  $E_{\gamma} > 1 MeV$, CalcHEP\cite{bib:calchep}\\
$e^+e^-\to e^+e^- \gamma$  & 188 mb &  $E_{\gamma} > 1 MeV$, CalcHEP\\
\hline
\end{tabular}
\label{tab:mmiss-bkg}
\end{table}

Three experiments, PADME at the DA$\Phi$NE Linac at LNF-INFN \cite{bib:padme-tdr}, 
MMAPS at Cornell \cite{bib:mmaps}, and at VEPP3 at Novosibirsks \cite{bib:vepp3}, 
are planning to exploit this technique. They share 
 similar design properties.  

\begin{figure}[htb]
    \begin{center}
        {\includegraphics[width=0.8\textwidth]{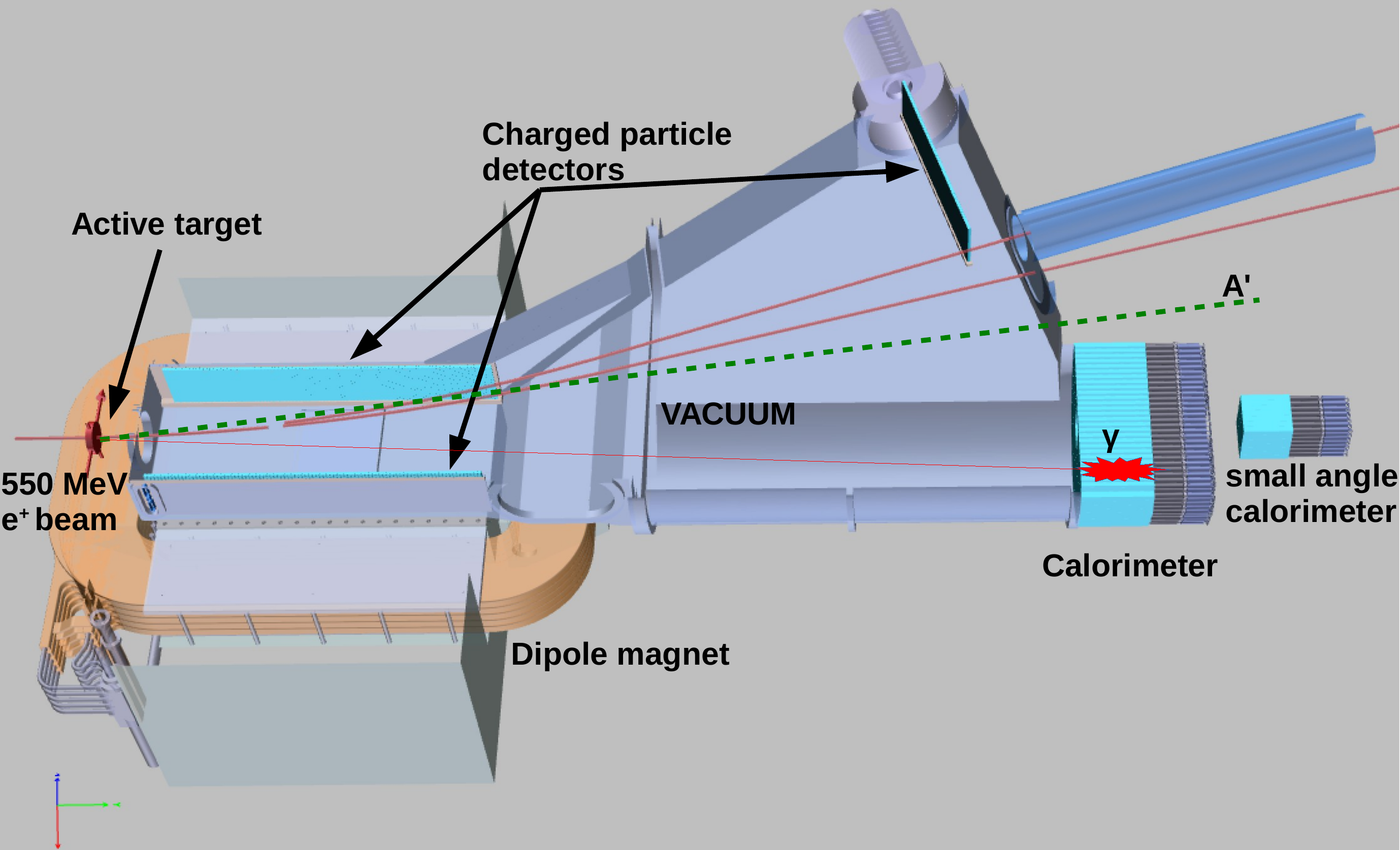}}
        \caption{\it CAD schematics of the PADME experiment at Frascati Linac.}
	\label{fig:padme}
    \end{center}
\end{figure}

PADME experiment at LNF-INFN, shown in fig. \ref{fig:padme}, will use 550 MeV positron beam
impinging on a 100 $\mu$m thick active target which is made of polycrystalline diamond. 
The recoil photons from the process  $e^+ + e^- \to \gamma A'$ 
will be detected by a ring-shaped BGO crystal calorimeter, which is located
3 m downstream and provides energy and position information. 
The non-interacted positron beam will be deflected outside the acceptance of the 
calorimeter by a dipole magnet. 
Three sets of plastic scintillator detectors will serve to register 
the charged particles and will provide an efficient veto 
for the bremsstrahlung background. 
In addition, a fast detector for photons, placed along the undeflected beam axis, 
will help with the suppression of the three photon annihilation background. 
The complete setup is located in vacuum to diminish the possible
beam-residual gas interactions. 

The approach of VEPP3 is to use an internal hydrogen gas target
placed at the Novosibirsks storage ring, illuminated by the 500 MeV $e^+$ beam.
Both the VEPP3 and the MMAPS experiments aim to use CsI crystals 
from CLEO for their calorimeters, which are shown to provide 
$\sigma_{E}/E = 3\%$ for 180 MeV positrons. 
In order to be able to operate in parallel with the ongoing 
VEPP3 activities an extension of the existing beam line is proposed.
This would allow to place a charged particle veto detector to 
suppress the background.
MMAPS uses a beryllium target and will
profit from the higher beam energy - up to 5.3 GeV. 
This extends MMAPS sensitivity region to masses up to $m_{A'}\sim 74$~MeV. 

\begin{table}[t]
\centering
\caption{ \it Comparison between the experiments exploiting the missing mass technique}
\vskip 0.1 in
\begin{tabular}{|l|c|c|c|} \hline
           &  PADME & MMAPS & VEPP3 \\
\hline
\hline
Place       &  LNF  &  Cornell &  Novosibirsk        \\
Beam energy  &  550 MeV & Up to 5.3 GeV &  500 MeV \\
$M_{A'}$ limit & 23 MeV &  74 MeV  & 22 MeV \\
Target thickness [e$^-$/cm$^2$] & $2\times10^{22}$& O($2\times10^{23}$) & $5\times10^{15}$ \\
Beam intensity & $8\times10^{-11}$ mA & $2.3\times10^{-6}$ mA & 30 mA \\
$e^+e^- \to \gamma\gamma$ rate [s$^{-1}$] & 15 & $2.2 \times 10^{6}$ & $1.5\times 10^{6}$ \\
$\epsilon^2$ limit (plateau) & $10^{-6}$  & $10^{-6}$ - $10^{-7}$ & $10^{-7}$ \\
Time scale   & 2017-2018 & ? & 2020 (ByPass) \\
Status & Approved & Not funded & Proposal \\
\hline
\end{tabular}
\label{tab:mmiss-cmp}
\end{table}

A comparison of the characteristics and the performance of the 
PADME, MMAPS, and VEPP3 experiments is shown in table \ref{tab:mmiss-cmp}.
PADME will be the first experiment to run and possible upgrades 
of the DA$\Phi$NE linac could extend its sensitivity down to 
the region of $\epsilon^2 \sim 3-5\times10^{-7}$. The VEPP3 experiment requires 
an approval of beam line modification while the Cornell MMAPS experiment 
is in process of identification of finding.

\subsection{Missing energy technique}

During the shower development inside a calorimeter,
the dark photons could be created through an A'-strahlung process
by any of the secondary particles. 
They could account for a large part 
of the undetected energy of the 
primary particle and their existence may 
manifest itself in the form of
a large fraction of events with missing energy.
This technique relies on the precise
knowledge of the response of a calorimeter to 
an electromagnetic shower development.  
It is exploited by the recently approved NA64 experiment
operating at CERN SPS \cite{bib:na64}. 
Synchrotron radiation 
tagged 100 GeV $e^-$ beam with momentum determined in a 
magnetic spectrometer impinges on an electromagnetic calorimeter (ECAL)
acting as an active beam dump. Events with less than 
50 GeV energy in the ECAL and no signal in the veto and in the 
downstream hadron calorimeter are considered as signal. 
In the hypothesis of zero background observation the NA64 experiment 
aims to set limits down to $\epsilon^2\sim 10^{-8} - 10^{-6}$ for A' masses 
in the region 10 MeV - 100 MeV, assuming $10^{10}$ electron events.

\subsection{Scattering of dark matter particles}
Another possible indication of the presence of
dark photons decaying into invisible particles $\chi$ could 
be the direct observation of the $\chi$ scattering in a 
low noise detector. 
This approach is similar to the beam dump technique widely used in the 
past to search for milicharged particles \cite{bib:tom-vulcano}. 
The A' are produced, through A'-strahlung
and then decay to $\chi\chi$ pair.
The BDX experiment \cite{bib:bdx}, proposed to take place at JLaB, 
will use a CsI (Tl) detector shielded by an active vetoing system 
to suppress the cosmogenic background. Studies indicated that 
 sensitivity down to $\epsilon^2\sim10^{-9}$ for $m_{A'}=50 $ MeV could be reached. 
However, this depends on the extra parameters $m_{\chi}$ and $\alpha_D$.

\section{Conclusions}

\begin{figure}[htb]
    \begin{center}
        {\includegraphics[width=0.7\textwidth]{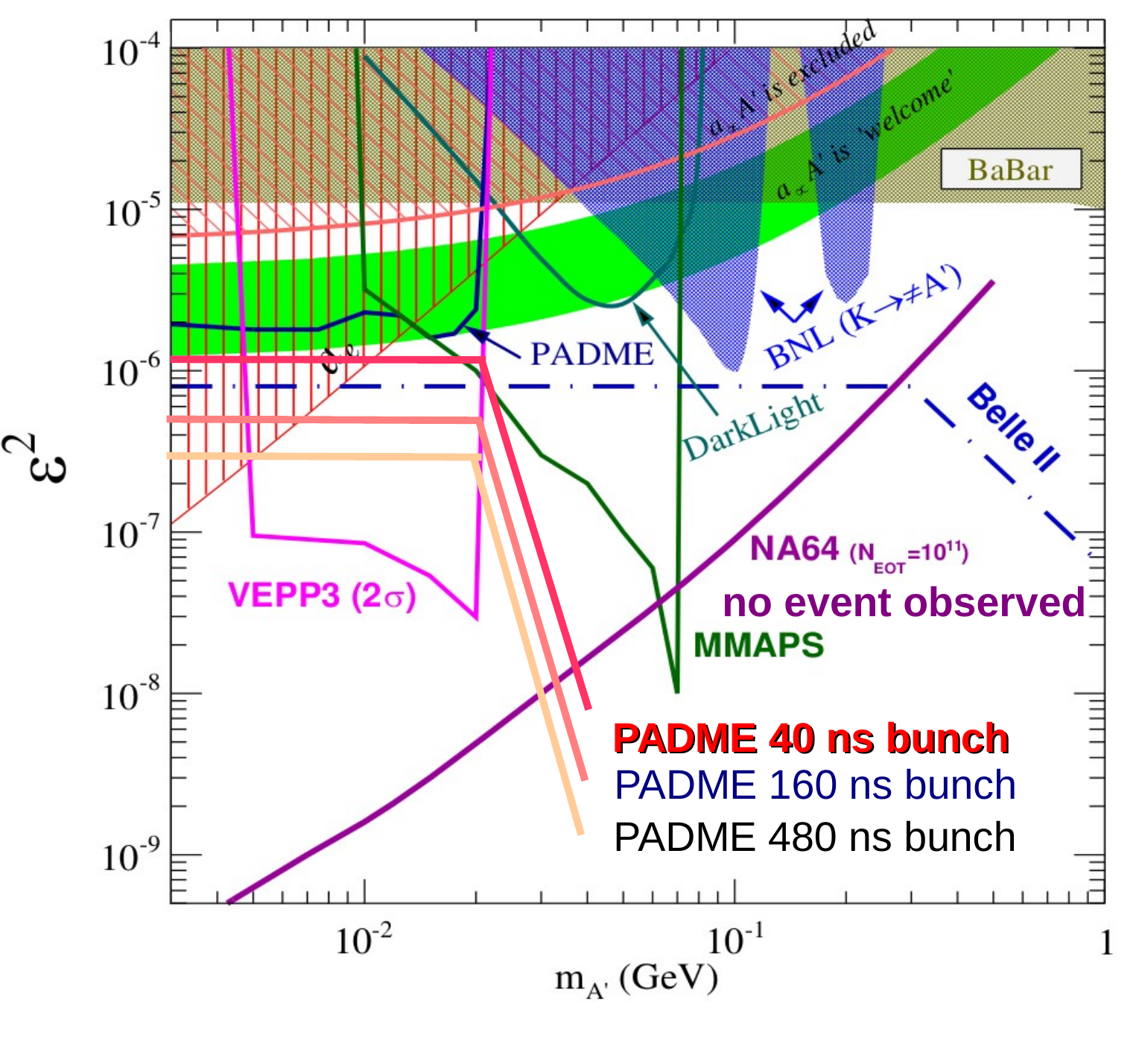}}
        \caption{\it Recent perspectives for invisible dark photon searches.}
	\label{fig:invisible}
    \end{center}
\end{figure}

Present searches for new vector particle A'
aim to cover 
the visible and the invisible final state scenarios in parallel. 
While on the visible side many results appeared in the 
last decade the parameter space 
for A' decaying predominantly to non Standard Model particles 
is being addressed just recently. 
The new activities in this direction can be summarized in fig. \ref{fig:invisible}. 
The PADME and the NA64 experiments are approved and are expected to provide 
interesting results by 2020. At the same time few other unique projects are in preparation. 


\section{Acknowledgements}
The authors would like to thank the organizers for the invitation to
present at the Vulcano workshop. VK acknowledges support from INFN,
LNF-INFN under contract LNF-SU 70-06-497/07-10-2014 
and University of Sofia science fund under grant FNI-SU N39/2016. 


%
\end{document}